\documentclass[a4paper,10pt,pra,aps,twocolumn,showpacs]{revtex4-1}

\usepackage{amsmath,amssymb,graphicx}
\usepackage{braket, datetime}

\newcommand{\figref}[1]{Fig.~\ref{#1}}
\newcommand{\eqnref}[1]{Eqn.~(\ref{#1})}
\newcommand{\secref}[1]{Section~\ref{#1}}

\newcommand{\rt}{\ensuremath{{\mathbf{r},t}}}
\newcommand{\rptp}{\ensuremath{{\mathbf{r'},t'}}}
\newcommand{\rrtt}{\ensuremath{{\mathbf{r},\mathbf{r'},t,t'}}}
\newcommand{\RRtt}{\ensuremath{{\mathbf{R},\mathbf{R'},\tau,\tau'}}}
\newcommand{\creation}{\ensuremath{\hat{\Psi}^\dagger(\rt)}}
\newcommand{\annihilation}{\ensuremath{\hat{\Psi}(\rt)}}
\newcommand{\creationp}{\ensuremath{\hat{\Psi}^\dagger(\rptp)}}
\newcommand{\annihilationp}{\ensuremath{\hat{\Psi}(\rptp)}}
\newcommand{\GOne}{\ensuremath{G^{(1)}}}
\newcommand{\GTwo}{\ensuremath{G^{(2)}}}
\newcommand{\gOne}{\ensuremath{g^{(1)}}}
\newcommand{\gTwo}{\ensuremath{g^{(2)}}}
\newcommand{\dr}{{\rm d}^3 {\bf r}}

\newcommand{\dt}{{\rm d}t}

\newcommand{\mean}[1]{\ensuremath{{\mathrm{E}\left({#1}\right)}}}
\newcommand{\variance}[1]{\ensuremath{{\mathrm{Var}\left({#1}\right)}}}
\newcommand{\figwidth}{0.45\textwidth}
\newcommand{\real}[1]{{\rm Re\left[ {#1}\right]}}

\begin{document}

\title{Temporal and spatio-temporal correlation functions for trapped Bose gases}
\author{M. Kohnen}\email[Current address: QUEST Institute, Physikalisch-Technische Bundesanstalt
and Leibniz Universit\"at Hannover, Bundesallee 100, 38116 Braunschweig, Germany]{} \
\author{R. A. Nyman}\email[Correspondence should be addressed to ]{r.nyman@imperial.ac.uk}
\affiliation{Centre for Cold Matter, Blackett Laboratory, Imperial College
London, Prince Consort Road, SW7 2BW, United Kingdom}
\date{\today}
\begin{abstract}
	Density correlations unambiguously reveal the quantum nature of matter. Here, we study correlations between measurements of density in cold-atom clouds at different times at one position, and also at two separated positions. We take into account the effects of finite-size and -duration measurements made by light beams passing through the atom cloud. We specialise to the case of Bose gases in harmonic traps above critical temperature, for weakly-perturbative measurements. For overlapping measurement regions, shot-noise correlations revive after a trap oscillation period. For non-overlapping regions, bosonic correlations dominate at long times, and propagate at finite speeds. Finally, we give a realistic measurement protocol for performing such experiments.
\end{abstract}
\pacs{03.75.Hh, 03.75.Kk, 67.85.Jk}
\maketitle

\section{Introduction}
Correlations between density measurements in quantum gases are related to the underlying correlation functions of the quantum field. In cold-atom experiments which detect individual atoms extracted from a cloud or a beam, the correlations appear in the arrival times and positions of the detected atoms\cite{Yasuda96,Schellekens05,Oettl05,Jeltes07,Guarrera11, Hodgman11, Guarrera12,Dall13}. Atoms are more (bosons) or less (fermions) likely to be coincident than classical statistics would suggest. 

By imaging a cloud after expansion and inspecting the fluctuations in the detected light, it is possible to infer correlations in momentum space\cite{Altman04}. This technique has been used to probe quantum many-body states such as the bosonic Mott insulator in a lattice\cite{Foelling05} and fermion pairing\cite{Greiner05}. Destructive \textit{in situ} imaging has revealed spatial correlation information about one-dimensional quasicondensates (through both pair\cite{Esteve06} and triple\cite{Armijo10} correlations) and Fermi gases\cite{Mueller10,Sanner10}. More sophisticated optical detection schemes are used to measure correlations in spin polarisation\cite{Sanner11, Meineke12} rather than number density.

Spatial density correlations can be inferred from fluctuations about the mean density\cite{Esteve06} . Fluctuations are related to temperature and to thermodynamic susceptibilities like compressibility via fluctuation-dissipation theorems\cite{Zhou11}. As susceptibilities tend to diverge close to phase transitions (e.g. Bose-Einstein condensation), so the measurable fluctuations become larger. However, fluctuation-dissipation theorems give information about the magnitude of fluctuations but not their length or time scales\cite{Pathria}. Naraschewski and Glauber\cite{Naraschewski99} showed how the underlying first-order correlation function relates straightforwardly to observable density-density\cite{Goldstein98} (second-order) correlations under the strong assumption that the system being measured is in the grand canonical ensemble at thermal equilibrium. In this paper we adapt their theory to include temporal correlations.

Many of the measurements of correlation functions in atomic gases were done by detecting atoms which had already been extracted from an expanding cloud or beam. Other experiments measured correlations between spatially separated regions of a cloud at one given instant or fluctuations from one experimental run to another. In the recent experiments of Guarrera \textit{et al.}\cite{Guarrera11, Guarrera12} atoms were extracted one by one from a single cloud and the  temporal correlations of their arrival times measured. There, the extraction of the first atom and its detection had no direct effect on the subsequent behaviour of the cloud: the correlations were present independently of the measurement, and the measurement itself was in some sense weakly perturbative\cite{Aharonov90}. 

In contrast to previous experiments, we consider here density-density (not atom-atom) correlation measurements using light beams which pass through a single atom cloud at different times. The most general version of the measurement includes sampling the density at differing positions and times. These retarded correlations have not yet been observed in cold atoms.

Correlations persist over thermal timescales like \mbox{$\hbar/ k_B T$} at temperature $T$. A typical length scale is the thermal de Broglie wavelength: $\lambda_T=\sqrt{2\pi\hbar^2 / M k_B T}$ for atoms of mass $M$. Taking a cloud of $^{87}$Rb at 200~nK the typical correlation times and lengths (without time-of-flight expansion) are 40~$\mu$s and 0.4~$\mu$m. Measurements can happen faster than 10~$\mu$s~\cite{Kohnen11b}, with spatial resolution down to about 1~$\mu$m, limited by photon collection and diffraction respectively. We therefore expect to be able to resolve correlations more easily in the time domain than in space.

In this article we first state our method for calculating correlation functions for thermal bosons and then show how optical measurements of atomic density relate to the correlation functions. We then specialise our discussion to the example of non-condensed Bose gases in harmonic traps. Finally, we discuss plausible experimental implementations using weakly-destructive measurements.

\section{Second-order correlation measures in the grand canonical ensemble}

Following the formalism of Ref.~\cite{Naraschewski99}, but retaining time dependence wherever possible, we start by defining the atomic-field annihilation operator for an atom at position ${\bf r}$ and time $t$ as \annihilation. Its Hermitian conjugate is the creation operator, and they follow the commutation relations for bosonic operators
\begin{align}
\left[ \annihilation, \annihilationp \right] &= 0 \\
\left[ \annihilation, \creationp \right] &= D({\bf r,r'}, t-t')
\label{eqn: commutation relation 2}
\end{align}
The function $D$ depends on the Hamiltonian governing the system (see Appendix~\ref{sec:unequal commutation}).

Density is an observable which is the expectation of the number density operator, \mbox{$n(\rt) = \braket{\hat{n}(\rt)}$} where {$\hat{n}(\rt) = \creation\annihilation$}. The quantum expectation value is denoted with angle brackets, and for readability we will use both notations $n(\rt)$ and $n_\rt$ for expectation of the density at a specified position and time.

We define normally-ordered, time-ordered ($t'\geq t$), first- and second-order atom-atom \textit{correlation functions}\cite{Naraschewski99, Goldstein98}:
\begin{align}
	\GOne_\rrtt&= \braket{\,\creation\annihilationp\,}\\
	\GTwo_\rrtt&= \braket{\,\creation\creationp \annihilationp\annihilation\,}
\end{align}
The first-order correlation function (with equal time arguments) \GOne\ is the spatial representation of the one-particle density operator. Also, the density and correlation function are simply related: \mbox{$n(\rt)=\GOne({\bf r, r},t,t)$}. The correlation between pairs of atoms is found in \GTwo. 

An obvious local, second-order, density \textit{correlation measure}, as distinct to an atom-atom correlation function, would be
$
	\braket{\hat{n}_\rt \hat{n}_\rptp} - \braket{\hat{n}_\rt}\braket{\hat{n}_\rptp} 
$
. However, the operator in the first term is not Hermitian, and so this measure is not a good observable. We conjecture that the appropriate observable is simply the real part.

We suppose that the measurements to be made are in some sense very weak, so that it does not matter in what order the measurements are made. The observable should then be symmetric under exchange of co-ordinates, $\set{{\bf r},t}$ and $\set{{\bf r}',t'}$. We note that when the variables are exchanged, the commutator in \eqnref{eqn: commutation relation 2} is complex conjugated, and thus the real part is a correct, symmetric operator. Therefore, the correlation measure that we will use is:
\begin{align}
	C_\rrtt=
		\frac{1}{2}\braket{\hat{n}_\rt \hat{n}_\rptp + \hat{n}_\rptp \hat{n}_\rt}
		- \braket{\hat{n}_\rt}\braket{\hat{n}_\rptp} \label{eqn:correlation estimator}
\end{align}
The purpose of defining this density correlation measure is to account for measurements over finite volumes and times, as will be shown in section \secref{sec:finite vol theory}.

Applying the definitions of \GOne and \GTwo, and the commutation relations, we note that ensemble average product of non-Hermitian operators can then be expressed in terms of the correlation functions:
\begin{align}
	\braket{\,\hat{n}_\rt \hat{n}_\rptp\,} = \GOne_{{\bf r,r'},t,t'}{D({\bf r,r'},t-t')} + \GTwo_\rrtt \label{eqn:density product expectation}
\end{align}

Since \GTwo\ is real:
\begin{align}
	\frac{1}{2}&\langle{\hat{n}_\rt \hat{n}_\rptp + \hat{n}_\rptp \hat{n}_\rt}\rangle
		= \\ 
	&\real{ \GOne_{{\bf r,r'},t,t'}{D({\bf r,r'},t-t')}}
	+ \GTwo_\rrtt \nonumber
\end{align}

For a non-interacting gas in a confining potential (with localised eigenfunctions) at thermal equilibrium, above the critical temperature for bosons, in the grand canonical ensemble, it can be shown that\cite{Naraschewski99} 
\begin{align}
	\GTwo_\rrtt = \GOne_{{\bf r,r},t,t} \GOne_{{\bf r',r'},t',t'} \, + \, \left|\GOne_\rrtt\right|^2
\end{align}
where the second term is due to bosonic exchange symmetry. The correlation measure can be expressed in terms of the first-order correlation function \GOne\ only:
\begin{align}
	C_\rrtt = \real{\GOne_{{\bf r,r'},t,t'}{D({\bf r,r'},t-t')}} \,+\, \left| \GOne_\rrtt\right|^2 \label{eqn:Crrtt}
\end{align}
The first term corresponds to counting (shot noise) number fluctuations. The second term comes from quantum exchange correlations.

\subsection{The first-order correlation function \GOne\ for non-interacting, trapped, thermal bosons}

\GOne\ is conveniently expressed in a basis of the eigenstates of the trapping potential. First we write the field creation and annihilation operators in this basis set:
\begin{align}
	\annihilation = \sum_{{\rm all}\,i} u_i({\bf r})e^{-{\rm i} \epsilon_i t / \hbar}\, \hat{a}_i \label{eqn:operator representation}
\end{align}
where the state labelled $i$ has energy $\epsilon_i$, spatial variation $u_i({\bf r})$ and annihilation operator $\hat{a}_i$. The sum runs over all states. The expectation $\braket{\hat{a}^\dagger_i \hat{a}_j}=\delta_{ij} N_i$, with $N_i$ being the thermal occupation number of the state, is given by the Bose-Einstein distribution, \mbox{$N_i = \zeta e^{-\beta \epsilon_i} / \left( 1 - \zeta e^{-\beta \epsilon_i}\right)$}. $\beta = 1/ k_B T$ is the inverse temperature, and $\zeta=e^{\beta \mu}$ the fugacity with $\mu$ being the chemical potential. It follows that the first-order correlation function is:
\begin{align}
	\GOne_\rrtt = \sum_{{\rm all}\,i}
			u^*_i({\bf r}) u_i({\bf r'}) e^{{\rm i} \epsilon_i (t-t') / \hbar}\,
				\frac{{\zeta} e^{-\beta {\epsilon}_i} }{1 - {\zeta} e^{-\beta {\epsilon}_i}}
	\label{eqn:G1 general form}
\end{align}

For bosons, it turns out to be helpful to expand the thermal occupation factor in \eqnref{eqn:G1 general form} as a power series of the fugacity, of the form $x/(1-x)=\sum_{k=1}x^k$. Combining exponentiated terms, we find:
\begin{align}
	\GOne_\rrtt \!\!=\!\! \sum_{k=1}^\infty {\zeta}^k  \sum_{{\rm all}\,i}
	\!u^*_i({\bf r}) u_i({\bf r'}) 
	e^{{\rm i} [(t-t') + {\rm i}k\hbar\beta]\epsilon_i/\hbar} \label{eqn:G1 general series expansion}
\end{align}
The inner sum is in the same form as the representation of the propagator, or Green's function\cite{Sakurai}, for the single-particle Hamiltonian $\hat{H}$:
\begin{align}
	K(\rrtt) =& \sum_{{\rm all}\,i} u^*_i({\bf r}) u_i({\bf r'}) e^{{\rm i}\epsilon_i (t-t')/\hbar} 
		\label{eqn: propagator general}\\
	=& \braket{{\bf r'}|e^{-{\rm i}\hat{H} (t'-t)/\hbar}|{\bf r}}\nonumber
\end{align}
Using this propagator with a complex time argument, the correlation function can be evaluated:
\begin{align}
	\GOne_\rrtt = \sum_{k=1}^\infty {\zeta}^k K({\bf r},{\bf r'},t,[t' - {\rm i}k\hbar\beta])\label{eqn:G1 from K}
\end{align}
At this point, we note that the expansion is effectively over increasing thermal occupation numbers, and that correlations with larger particle numbers will decay rapidly for non-condensed systems. 

The fugacity can be found by integrating the density over all space to give the total atom number:
\begin{align}
	N_{total} &= \int \dr \,\GOne_{{\bf r,r},t,t} \label{eqn:fugacity constraint}
\end{align}
If the atom number is to be fixed in simulation, then the fugacity must be found so that this equation is satisfied.

\subsection{Correlation measurements over finite volumes}\label{sec:finite vol theory}

Imaging experiments measure the change of amplitude, phase or polarisation of a probe light beam due to the atom-light interaction\cite{Goldstein98}. In the limit of small optical density, the change to the light beam is proportional to the average atomic density seen by the beam as it passes through the cloud, necessarily integrating along line of sight. There will also be a finite resolution transverse to the beam due to the optical geometry, ultimately limited by diffraction. Therefore, any optical measurement of density covers a finite volume. Real measurements also take finite times to count sufficient photons. This finite detection volume and time will tend to smear out and reduce measured correlations.

The expectation of the time-averaged number of atoms in the probe beam during a measurement is\footnote{The notation \mean{X} for the expectation value of a random variable is fairly standard when distinguishing between stochastic processes (e.g. experimental realisations) and quantum expectation values\cite{BreuerPetruccione,WuertzThesis}. The interpretation of the two kinds of expectation, statistical and quantum, is of course rather tricky, but here we are just making use of the  clarity of notation.}:
\begin{align}
	\mean{N_{{\bf R},\tau}}=\int \dt \int \dr\,\, \braket{\hat{n}_\rt} I_{{\bf R}\tau}({\bf r},t)
	\label{eqn:Nrt}
\end{align}
with ${\bf R}$ a label for the central position of the probe beam for measurements around time $\tau$. The probe beam has a spatial and temporal intensity profile $I_{{\bf R},\tau}({\bf r},t)$ which is normalised such that $\int \dt \,I_{{\bf R},\tau}({\bf R},t) = 1$ (dimensionless) at its spatial maximum\cite{Nyman11}. Photons arriving at detectors will be used to infer this atom number.

\begin{figure}[hbt]
	\centering
	\includegraphics[width=0.45\textwidth]{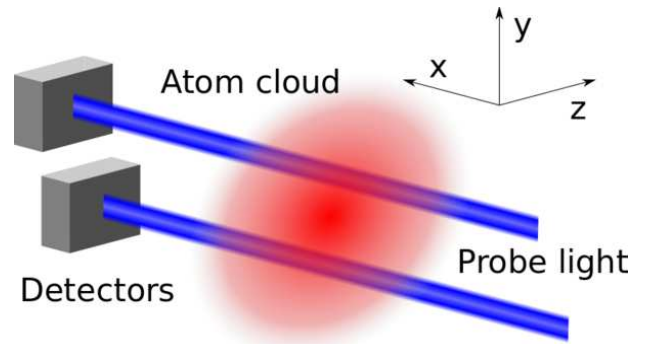}
	\caption{A correlation measurement requires at least two atom number measurements at different times and places. Light beams pass through the atom cloud, and the number of atoms in the interaction volume shows up in the amplitude or phase properties of the light.}
	\label{fig:diagram}
\end{figure}

A general second-order correlation measurement includes a series of paired number measurements at two times, $\tau$ and $\tau'$ and with probe beams in modes centred at different positions ${\bf R}$ and ${\bf R}'$, as shown schematically in \figref{fig:diagram}. This measurement corresponds to measuring the product of densities, which is the only kind of correlation that can be measured using near-resonant light\cite{Goldstein98}. A good estimator for the correlations, the covariance, noted with the tilde to remind us that it's spatially integrated and averaged over a series of realisations, is:
\begin{align}
   \tilde{C}&_\RRtt \nonumber\\
   &=
	\iint \dt \,\dt'
	\iint \dr\,\dr'
	I_{{\bf R},\tau}({\bf r},t) I_{{\bf R'},\tau'}({\bf r'},t')
	\\
	&\hspace{8ex}
	\,
	\left[
	\frac{1}{2}\braket{\hat{n}_\rt \hat{n}_\rptp + \hat{n}_\rptp \hat{n}_\rt}
		- \braket{\hat{n}_\rt}\braket{\hat{n}_\rptp}
	\right] \nonumber
\end{align}
This finite-volume correlation estimator is the double integral of the local correlation measure, \eqnref{eqn:correlation estimator}, weighted by the probe beam intensities.

This can in turn be expressed in terms of \GOne, as in \eqnref{eqn:Crrtt}. We find:
\begin{align}
   \!\!\tilde{C}_\RRtt &\!=\! \label{eqn:correlation estimator finite}
	\iint \!\dt \,\dt'
	\iint \!\dr\,\dr'
	\,I_{{\bf R},\tau}({\bf r},t) I_{{\bf R'},\tau'}({\bf r'},t')\times\nonumber\\
	&\hspace{-5ex}
		\left\{\real{\GOne_{{\bf r,r'},t,t'} D({\bf r,r'},t-t')} \,+\, \left| \GOne_\rrtt\right|^2\right\}.
\end{align}

It is common to work with correlation estimators which are normalised by the standard deviations of each of the component measurements. In this case, we choose to approximate the standard deviations of measured numbers with Poissonian statistics in the limit of large number: $\variance{N}= \mean{N}$, where \variance{X} is the sample variance. The normalised correlation estimator is then:
\begin{align}
	\tilde{c}_\RRtt = \frac{\tilde{C}_\RRtt }{ \sqrt{\mean{N_{{\bf R},\tau}}\mean{N_{{\bf R'},\tau'}}}}
\label{eqn:normalised_correlation_estimator}
\end{align}

\section{Localised correlations in Bose gases in harmonic traps}
We now turn to the specific case of thermal bosons in a harmonic trap. Our task is to evaluate the local correlation estimator, $C_\rrtt$. The Hamiltonian for an atom of mass $M$ in a trap of frequencies $\Omega_x, \Omega_y$ and $\Omega_z$ is \mbox{$\hat{H}=-(\hbar^2/ 2M)\nabla^2 + M(\Omega_x^2 x^2 + \Omega_y^2 y^2 + \Omega_z^2 z^2)/2$}. The eigenstates $u_{lmn}({\bf r})$ are the Hermite polynomials with energies \mbox{$\epsilon_{lmn}=\hbar\left(l\Omega_x+m\Omega_y+n\Omega_z\right)$}, with zero-point energy subtracted for ease of notation. Since the Hamiltonian is a sum of three terms, one for each dimension, both the eigenstates and the propagator are separable: \mbox{$K_\rrtt = K_x(x,x',t,t')K_y(y,y',t,t')K_z(z,z',t,t')$}. 

The propagator along the $x$ direction, for the $k$th term in the expansion of the correlation function, is given by\cite{Sakurai}: 
\begin{align}
 K&^{(k)}_x(x,x',t,[t'-{\rm i}\hbar k\beta])= \label{eqn: propagator}\\
	&\sqrt{
	\frac{M\Omega_x}
		{2\pi{\rm i} \hbar \sin{[\Omega_x(t'\!-\!t\!-\!{\rm i}k\hbar\beta)]}}} 
	\,\,\, \times\nonumber\\
	&\hspace{1ex} \exp
	{\left[ 
\frac{{\rm i}M\Omega_x \left\{ (x'^2+x^2)\cos[\Omega_x(t'\!-\!t\!-\!{\rm i}k\hbar\beta)] -2x'x\right\}}{2\hbar\sin[\Omega_x(t'\!-\!t\!-\!{\rm i}k\hbar\beta)]}
\right]}
	\nonumber
\end{align}
With a real argument this propagator oscillates, so there will be damped oscillations in the correlation function from the complex argument.

The correlation function is then a sum over products of partial propagators:
\begin{align}
	\GOne_\rrtt = \sum_{k=1}^\infty {\zeta}^k \prod_{s=x,y,z} 
			K^{(k)}_s&(s,s',t,[t'-{\rm i}\hbar k\beta])
	\label{eqn:G1 from KxKyKz}
\end{align}
Note that this does not mean that the correlation function is separable.

By numerical evaluation of the fugacity constraint \eqnref{eqn:fugacity constraint}:
\begin{align}
	N_{total} &= \sum_{k=1}^\infty {\zeta}^k \!\!
		\prod_{s=x,y,z}
		\int_{\rm all\,space} \hspace{-5ex}{\rm d}s \, K_s^{(k)}(s,s,t,[t - {\rm i}k\hbar\beta])
\end{align}
it is possible to choose the atom number, and then to iteratively estimate the fugacity $\zeta$ to make this equation consistent.

\subsection{Two-position correlations at one time}
\begin{figure}[hbt]
	\centering
	\includegraphics[width=\figwidth]{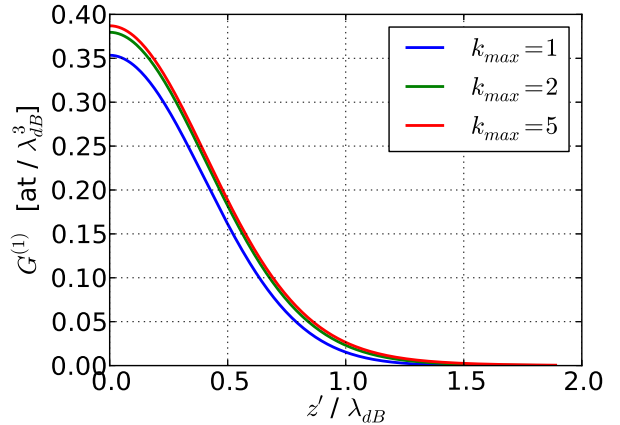}
	\caption{Un-normalised spatial correlation function \GOne, for a cloud of 20\,000~atoms of $^{87}$Rb at 500~nK in a trap of \mbox{$1000\times1000\times20$~Hz}, along the centre line of the trap at the time zero. Since the phase-space density is 0.35, large particle number occupancies ($k$) are not expected, and the series \eqnref{eqn:G1 from KxKyKz}  from $k=1$ to $k=k_{max}$ converges rapidly. The fugacity which gives the correct particle number is $\zeta=0.34$. The typical correlation length is the de Broglie wavelength, which is around 265~nm in this case.}
	\label{fig:convergence kmax}
\end{figure}
We now inspect the behaviour of the first-order correlation function \GOne\ for a cloud of 20\,000~atoms of $^{87}$Rb at 500~nK in a cigar-shaped trap of \mbox{$1\,\rm{kHz}\times 1\,\rm{kHz}\times 20$~Hz}, along the centre of the trap at time zero. In \figref{fig:convergence kmax}, we see that the series \eqnref{eqn:G1 general series expansion} converges rapidly for temperatures significantly above critical. Each term in $\zeta^k$ corresponds to a thermal occupation number of the ground state by $k$ particles. Since the temperature is above critical for this example, the probability of thermal occupation of large numbers of particles in the ground state is very small. The correlation function is significant over length scales around the de Broglie wavelength. 

For further discussion, we define the normalised correlation functions:\
\begin{align}
	\gOne_\rrtt &= \frac{\GOne_\rrtt }{ \sqrt{\GOne_{{\bf r, r},t,t}\GOne_{{\bf r', r'},t',t'}}}\\
	\gTwo_\rrtt &= \frac{\GTwo_\rrtt} {\GOne_{{\bf r, r},t,t}\GOne_{{\bf r', r'},t',t'}},
\end{align}
which are independent of the number of particles in the trap and start at $\gOne = 1$ and $\gTwo=2$ for $t=t'=0$ and ${\bf r}={\bf r'}$.

\subsection{Two-time correlations at one position}
\begin{figure}[htb]
	\centering
	\includegraphics[width=\figwidth]{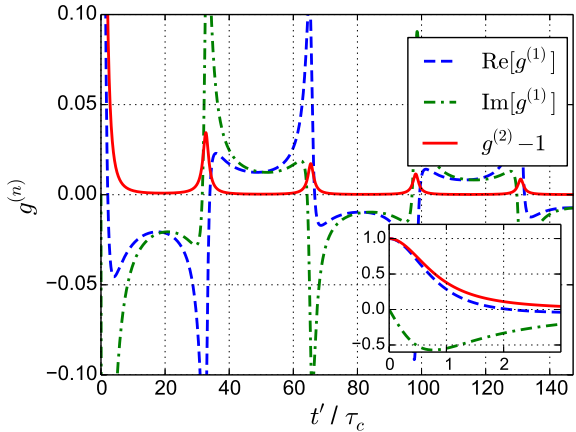}
	\caption{Normalised two-time correlation functions \gOne\ (real and imaginary parts) and $\gTwo-1$ for the same atom cloud as \figref{fig:convergence kmax}, at the centre of the trap. Inset: same figure but at shorter times. The correlation function decays on the thermal time scale $\tau_c = \hbar\beta=15~\mu$s. Correlations revive after one half oscillation of atoms along the tighter axis, 500~$\mu$s$=32.7 \times\tau_c$ here.}
	\label{fig:two time local}
\end{figure}

Turning our attention to the normalised, two-time correlation function, \figref{fig:two time local} shows a decay of the two-particle correlations on the time scale $\tau_c = \hbar\beta$ (inset). The correlations revive and decay as particles oscillate in the confining potential, returning partially after the half the shortest oscillation period, i.e 500~$\mu$s. The correlations return completely on the longest time scale in the system, half the axial oscillation period (25~ms in this case). Any two particles will separate and then return to their original position after half an oscillation in the trap, so the two-particle correlation function revives. The real part of \gOne\ is negative for most of the time between the beginning and the first revival. This phase shift could in principle be observable in an two-path interferometer based on extracting atoms from the cloud, and then relatively delaying one of the paths. Beats between the different trap frequencies appear in the correlation function.

\begin{figure}[hbt]
	\centering
	\includegraphics[width=\figwidth]{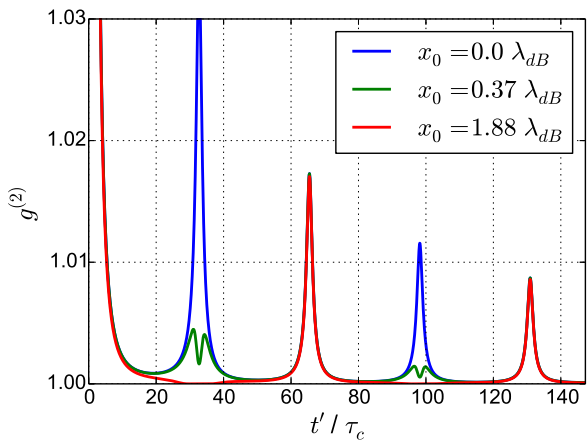}
	\caption{Normalised two-time correlation function $\gTwo$ for the same atom cloud as \figref{fig:convergence kmax}, but offset from the trap centre along the short direction by a distance $x_0$. We see that the revival goes from twice per oscillation period to just once as position is moved away from the trap centre.}
	\label{fig:two time local offset}
\end{figure}

Inspecting \gTwo\ at a fixed position but off-centre of the trap, \figref{fig:two time local offset} shows that the revival changes from occurring twice per oscillation period to just once. The distance off-axis required for this effect is approximately the de~Broglie wavelength.

\subsection{Two-time, two-position correlations}

\begin{figure}[ht]
	\centering
	\includegraphics[width=\figwidth]{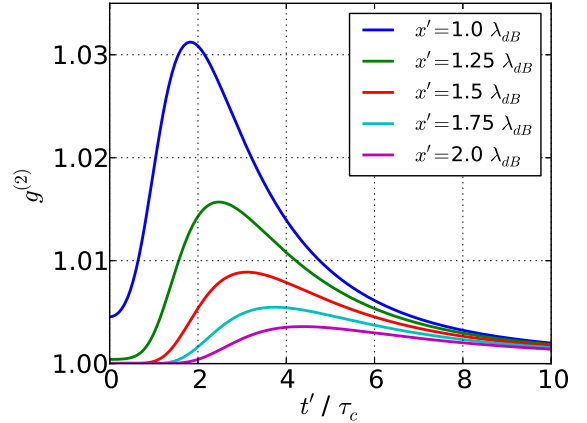}
	\caption{Normalised two-time correlation function $\gTwo$ for the same atom cloud as \figref{fig:convergence kmax}, as a function of both time $t'$ and position, with first point at the centre of the trap and time $t=0$, the second off centre at $x'$. We see how correlations take a finite time to reach the second position.}
	\label{fig:retarded local}
\end{figure}

\figref{fig:retarded local} shows the temporal evolution of $\gTwo$ when the measurement at the second time $\tau^\prime$ is done away from the trap centre. The correlation function peaks at later times at points further away from the centre. We can therefore define a speed at which the peak of the correlation function travels. It can be calculated by combining the typical correlation time scale $\tau_c=\hbar \beta$ with the de Broglie wavelength $\lambda_T$ to make the correlation velocity $v_c=\sqrt{2\pi k_B T/ M}$. We see from figure \figref{fig:retarded local} that correlations travel at about $2.5 \times v_c$.

\section{Measurements over finite volumes in oblate harmonic traps}

We now apply the general form of \secref{sec:finite vol theory} to the specific case of harmonically trapped Bose gases probed by Gaussian-shaped beams. The co-ordinate system is defined in \figref{fig:diagram}. We make the approximation that the Rayleigh range of the probe beam is much longer than the thickness of the cloud in the direction of propagation. This approximation makes it possible to separate the detection response function,  into components $I_{{\rm R},\tau}({\bf r},t)= I_X(x)I_Y(y)I_Z(z)I_\tau(t)$. 

Measurements can be performed quickly compared to the typical correlation time scales, $\hbar\beta$, so we will approximate the temporal response function of the detection with an infinitesimal response time: 
\mbox{
$I_{{\bf R},\tau}({\bf r},t)=I_{{\bf R}}({\bf r)}\delta(t-\tau)$.
}
The separated factors of the spatial detector response functions are $I_X(x)=1$, \mbox{$I_Y(y)=e^{-2(y-Y)^2/w_0^2}$}, and \mbox{$I_Z(z)=e^{-2(z-Z)^2/w_0^2}$}, where ${\bf R}=(X,Y,Z)$ and $w_0$ is the waist of the probe beam. 

We consider a pancake shaped atomic cloud, which is thin along the line of sight in order to minimise the loss of coherence through integration along the probe axis; the other axes are kept loose. The trap contains $10^5$~atoms at 100~nK and has trap frequencies of $413\times 20 \times 9$~Hz, much as in Ref. \cite{Guarrera11}.

The full expression for the correlation estimator with a finite detection volume, \eqnref{eqn:correlation estimator finite}, requires a numerically intractable 8-dimensional integral. Since the detector response contains a $\delta$-function we can execute the time integrals. The separability of both propagators and detector response can be used to express the 6-dimensional spatial integral as a product of two-dimensional integrals, reducing the numerical complexity.

%

Making use of the result of \eqnref{eqn: propagator to commutator}, we can now numerically evaluate \eqnref{eqn:correlation estimator finite} as:
\begin{widetext}
\begin{align}
\label{eqn:full correlation estimator}
	\tilde{C}_\RRtt =
	&
	\left\{
	\sum_{k=1}^\infty {\zeta}^k
	\real{
		\prod_{s=x,y,z} \iint \,{\rm d}s\,{\rm d}s'\, 
			I_{{S}}({s}) I_{{S'}}({s'})
				K^{(k)}_s(s,s',\tau,[\tau'-{\rm i}\hbar k\beta]) 
				K^{(0)}(s',s,\tau',\tau)
	}
	\right\} 
	\hspace{4ex}+\hspace{4ex}
	\\
	&\hspace{6ex}
	\left\{
	\sum_{k=1}^\infty \sum_{l=1}^\infty \prod_{s=x,y,z} \!\! {z}^{k+l} 
		\iint \,{\rm d}s \, {\rm d}s'\, 
			I_{{S}}(s) I_{{S'}}({s'})
			{K^{(k)}_s}^* (s,s',\tau,[\tau'-{\rm i}\hbar k\beta])
			K^{(l)}_s(s,s',\tau,[\tau'-{\rm i}\hbar l\beta])
	\right\}
	\nonumber
\end{align}
\end{widetext}

The first term is due to quantisation of the matter field into atoms (shot noise). This term is dominant around the peaks of $D(\mathbf r,\mathbf r', \tau^\prime - \tau)$ (see Appendix~\ref{sec:unequal commutation}). It is maximised at $\tau^\prime = \tau = 0$, where it reduces to 
\begin{align}
\label{eqn:sn}
\tilde{C}^{sn}_{\mathbf R, \mathbf R'} =& \sum_{k=1}^\infty {\zeta}^k \prod_{s=x,y,z} \int \,{\rm d}s\\ &I_{{S}}({s}) I_{{S'}}({s}) K^{(k)}_s(s,s,0,[-{\rm i}\hbar k\beta]). \nonumber
\end{align}
This equation gives an upper bound for the shot noise contributions. The second term in \eqnref{eqn:full correlation estimator} is due to bosonic fluctuations and will be labelled $\tilde{C}^{bs}$. It decays more slowly than the first term when increasing $\tau^\prime - \tau$.

\subsection{Two-position correlations at one time}

\begin{figure}[hbt]
	\centering
	\includegraphics[width=\figwidth]{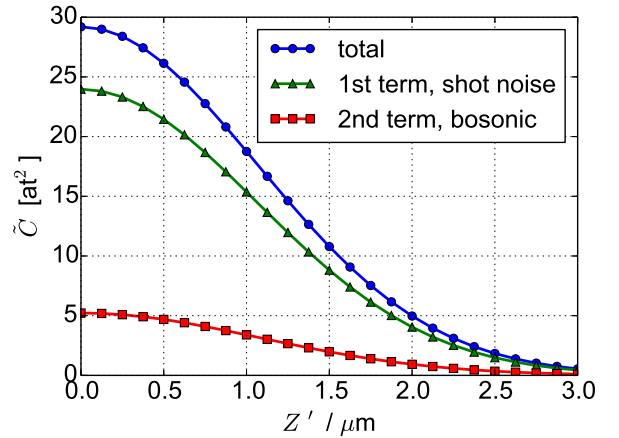}
	\caption{Un-normalised fluctuations $\tilde{C}$, for a cloud of $10^5$~atoms at 100~nK in a pancake-shape trap of $413\times 20 \times 9$~Hz as in Ref. \cite{Guarrera11}, for two measurements at nearly equal times (just 10~$\mu$s apart). The phase-space density is 0.82. The scale over which correlations decay is the size of the probe beam used for detection, which is 1.5~$\mu$m in this case. For comparison, the de Broglie wavelength is 0.59~$\mu$m. The maximum shot noise fluctuation calculated with \eqnref{eqn:sn} is 24 atoms. The mean atom number in the probe beam is 37, indicating that the fluctuations are effectively suppressed by the presence of atoms in the Gaussian wings of the detection volume.}
	\label{fig:two position measure}
\end{figure}

The correlation measure taken for two short (compared to $\hbar\beta$) measurements of the number of atoms in the probe beam, at nearly equal times, but two positions is shown in \figref{fig:two position measure}. The number of atoms in the beam estimated from equation \eqnref{eqn:Nrt} is 37 atoms. The measures are taken 10~$\mu$s apart since the numerical integrals do not converge for exactly equal times. The bosonic-enhancement of fluctuations is up to 5 atoms. The bosonic correlations decay on the length scale of the measurement probe beam size, in this case 1.5~$\mu$m. The bosonic-correlation length scale intrinsic to the atom cloud is $\lambda_T=0.59$~$\mu$m here, much smaller than beam size.  The shot-noise fluctuation at the origin is 24 atoms and therefore less than would be expected from Poissonian statistics, where variance equals the mean. The total mean atom number in the probe region is made of many atoms in the wings of the Gaussian probe beams, which are only partially detected, and hence fluctuations are reduced.

\subsection{Correlations at two times}

\begin{figure}[hbt]
	\centering
	\includegraphics[width=\figwidth]{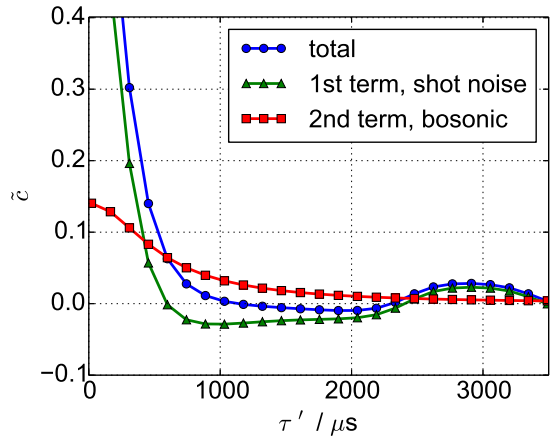}
	\caption{Normalised correlation estimator for the same cloud as \figref{fig:two position measure}, for two measurements at one position. The time scale over which correlations decay is the time a typical excitation takes to cross the probe beam used for detection, $\propto 1/\sqrt{T}$, and not the typical thermal timescale $\hbar \beta=76\,\mu$s. At short times, shot noise (single-particle autocorrelation) is the dominant contribution to the correlation measure. Shot noise also exhibits a revival at about one oscillation period of the tightest axis of the trap. }
	\label{fig:two time measure one temperature}
\end{figure}

We now turn to normalised correlation estimators at two times as defined in \eqnref{eqn:normalised_correlation_estimator}. \figref{fig:two time measure one temperature} shows how the one-particle autocorrelation [first part of \eqnref{eqn:full correlation estimator}] decays faster than the bosonic correlations. Furthermore, the shot noise demonstrates anticorrelations as particles which start in the middle move away. When they come back to the centre, a revival can be seen in the correlation function, roughly one full oscillation period along the tightest trap axis.

We also note that fluctuations become weaker as the temperature becomes much higher than the critical temperature for Bose-Einstein condensation. Furthermore, the time scale on which the correlations decrease is dictated by the time it takes atoms with the mean thermal speed to cross the detection region (again 1.5~$\mu$m here). Since this speed increases with temperature, so the correlations die away faster at higher temperatures, with a typical timescale proportional to $1/\sqrt{T}$ and not the thermal coherence time scale, $\propto 1/T$.

\begin{figure}[hbt]
	\centering
	\includegraphics[width=\figwidth]{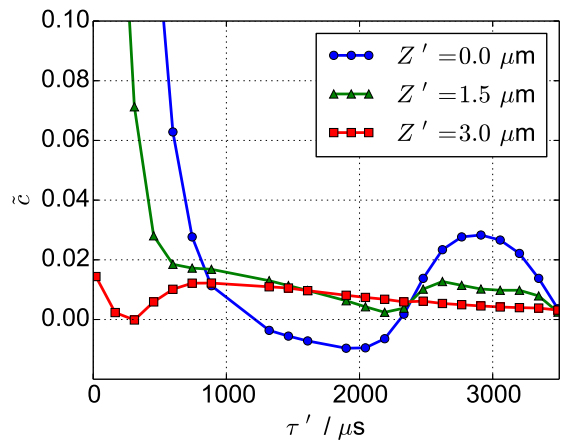}
	\caption{Normalised correlation estimator for the same cloud as \figref{fig:two position measure} for two measurements at different positions as a function of the delay between measurements, $\tau'$. For the furthest separated measurement positions, the bosonic correlations propagate at finite speed. Negative autocorrelations cause the initial dip, and revivals are not observed. For comparison, $\hbar \beta = 76\,\mu$s and $1/\Omega_x=2420\,\mu$s.
	}
	\label{fig:two time two position measure}
\end{figure}

When the two measurements at different times are also done at different positions, we see again that correlations take a finite time to move from the first point of measurement to the second. As shown in \figref{fig:two time two position measure}, when the probe beams are well separated ($Z^\prime > w_0$), the maximum of correlation is found at finite (not zero) time. This is an observable manifestation of the propagation of correlations at finite speed. For well-separated beams, the revival in correlations is not detectable. 

\section{Realistic measurement protocols and uncertainties}

The determination of atom number will come with some unavoidable, fundamental uncertainties. We will now estimate how many measurements it will take to distinguish the bosonic part of the correlations from the shot noise.

The uncertainty in the correlation measure, $\variance{\tilde{C}}$, assuming approximately Gaussian statistics, is\cite{Barlow}:
\begin{align}
	\variance{\tilde{C}} \simeq \frac{1}{{N_{runs}-1}} \variance{N_{{\bf R},\tau}}\,\variance{N_{\bf R',\tau'}}
\label{eqn:variance C}	
\end{align}
where $\variance{N_{{\bf R},\tau}}$ is the uncertainty in the measured number of atoms in the probe beam at position ${\bf R}$ and time $\tau$, and $N_{runs}$ is the number of experimental runs.

Optical detection of atoms always comes with incoherent photon scattering\cite{Lye03}, no matter what property of the light beam is measured. The more photons spontaneously emitted, the more precise the measurement can be. However, spontaneous emission destroys the second-order coherence that we want to measure. We therefore have to limit the precision of each measurement, in favour of more, independent, experimental runs to reduce the statistical uncertainties. The measurement uncertainty per spontaneously emitted photon can be decreased by using small probe beams, so it pays to use high numerical aperture optics. In a shot-noise limited measurement with a Gaussian beam of size $w_0$ the uncertainty in the inferred atom number is\cite{Nyman11}:
\begin{align}
	\variance{N_{{\bf R},\tau}} = {\frac{\pi w_0^2}{2 \sigma_0 n_{sc}}}
	\label{eqn:sigma N}
\end{align}
where $n_{sc}$ is the number of spontaneous emission events per atom and $\sigma_0$ is the resonant atom-light scattering cross section, which can be up to $3\lambda^2 / 2\pi$ for $\lambda$ the resonant optical wavelength. Note that this formula is only valid if the transverse size of the probe beam is significantly smaller than the atom cloud.

To detect the bosonic enhancement of the correlations, the variance on the correlations measure has to be less than the bosonic part of the correlation measure, the second term in \eqnref{eqn:full correlation estimator}:
\begin{align}
\variance{\tilde{C}} \leq \tilde{C}^{bs}
\end{align}
Using \eqnref{eqn:variance C} and \eqnref{eqn:sigma N} we get an expression for the number of measurements required to measure the bosonic part of the correlation measure with a signal-to-noise ratio of one:
\begin{equation}
	N_{runs}[SNR=1] \approx \left( \frac{\pi w_0^2}{2 \sigma_0 n_{sc}} \right)^2 \, \frac{1}{\tilde{C}^{bs}}
\end{equation}

Taking the example of Ref. \cite{Guarrera11} for experimental parameters, using a 1.5~$\mu$m Gaussian-shape probe laser beam and limiting the number of photons scattered to $0.2$ per atom, we find that a signal-to-noise ratio of 1 is achieved for about 750 experimental runs, for measurements at nearly equal positions and times. The cloud extends over a few times the size of the detection region, so it is possible to measure correlations for several pairs of probe beams in parallel, pairs being separated by much more than intra-pair spacing. Several measurements can be made on one cloud, although atom number will reduce and temperature will rise.

Only certain kinds of optical measurements are suitable. The optical densities (exponents in the Beer-Lambert law) of the Bose gas clouds discussed are typically larger than 1, so the detector response is not linear in atom number and \eqnref{eqn:Nrt} is not valid. Phase shift measurements (far off resonance) have a linear response, but will induce a net, localised, mean phase shift on the atoms due to the AC Stark effect. Such a phase shift will lead to mechanical effects on the atom cloud, such as heating or sound waves, which might disguise the atom-atom correlation signals. We have already developed a suitable two-frequency measurement technique which cancels this mean phase shift\cite{Kohnen11b}.

\section{Conclusions}

We have shown that the results of temporal and spatio-temporal correlation measurements in non-interacting thermal Bose gases reflect the underlying atom-atom correlation functions. We have explained how to measure these quantities optically, for Bose gases in harmonic traps in thermal equilibrium. Correlations unsurprisingly take time to travel finite distances. The revivals in the correlation functions are in practice probably too weak to be visible in finite-volume correlation measurements, especially in the presence of non-zero atom-atom scattering~\cite{Bezett14}. 

The two-time correlation functions can, in principle, be calculated in other quantum gas systems such as Bose-Einstein condensates, interacting Bose gases, Fermi gases, and for other kinds of potentials, e.g. box potentials or optical lattices.  Correlation measures for these systems would provide unambiguous evidence of exotic phase transitions.

One can expect that strong, projective measurements will take the ensemble out of thermal equilibrium. Extending the theory beyond thermal equilibrium is possible using, for example, stochastic time-evolution techniques\cite{Wilson07}. We would be very interested to read about the results of any such calculations.

\subsection*{Acknowledgements}

We have had some very enlightening discussions, especially about \eqnref{eqn:correlation estimator} to \eqnref{eqn:density product expectation} with, in alphabetical order: S.Y.~Buhmann, E.-M.~Graefe, E.A.~Hinds, M.S.~Kim, A.~Nazir, A.D.K.~Plato, A.~Sanpera and S.~Scheel. This work was supported by EPSRC grant EP/J017027/1.

\appendix

\section{Properties of the commutation relation at unequal times}\label{sec:unequal commutation}

In \eqnref{eqn:operator representation} we decompose the field operators into a basis set of the eigenstates of the Hamiltonian that governs the system. The commutation relations for creation and annihilation of particles within specific states are:
\begin{align}
	\left[\hat{a}_i, \hat{a}_j^\dagger\right]&=\delta_{ij}\\
	\left[\hat{a}_i, \hat{a}_j \right]&=0
\end{align}
where the $\delta_{ij}$ is the well-defined Kronecker delta. Explicitly writing out the creation-annihilation operator for the fields we find:
\begin{align}
	[\, \annihilation, \creationp\,] =& 
	\sum_{{\rm all}\,i,j} u_i({\bf r}) u^*_j({\bf r'})\, 
		e^{+{\rm i} (\epsilon_j t' -\epsilon_i t) / \hbar}
		[\hat{a}_i , \hat{a}^\dagger_j ]\nonumber\\
	=&\sum_{{\rm all}\,i} u_i({\bf r}) u^*_i({\bf r'}) e^{+{\rm i} \epsilon_i (t'-t)  / \hbar}
	\nonumber\\
	= & D({\bf r},{\bf r'}, t-t')
	\label{eqn: represented commutation relation}
\end{align}

With this formula we can calculate the unequal-time commutation relations for any bound (discrete) system, and the formula can be generalised to unbound (continuum) cases. By inspection, the commutator is directly related to the propagator of \eqnref{eqn: propagator general}. 
\begin{align}
	D({\bf r},{\bf r'}, t-t') = K({\bf r'},{\bf r}, t',t)\label{eqn: propagator to commutator}
\end{align}
At equal times \eqnref{eqn: represented commutation relation} reduces to the completeness relation for eigenfunctions, which is guaranteed by the fact that the Hamiltonian operator is Hermitian. 

At unequal times, the result depends on the Hamiltonian. For our case of a 3D harmonic oscillator, which is trivially separable, there exists an analytic formula for the propagator, \eqnref{eqn: propagator}. This shows that the commutator is non-zero even for unequal positions. A physical picture, for bound states, is that an initial delta-function wavepacket spreads out over time. Starting with two such wavepackets, with different positions at different times, there will be some overlap between them due to this spreading. Hence, their field operators do not commute, except for special cases like revivals in a harmonic oscillator. In those cases, the commutator tends to a Dirac delta function function of position in the limit where either the time arguments are equal or separated a multiple of the characteristic oscillation period.

In general, a measurement done at time $t$ will modify the state of the system and therefore change \mbox{$D({\bf r},{\bf r'}, t-t')$}. For the purpose of this article, however, we made the strong assumption that the measurement is only weakly perturbative and can therefore be neglected.

We also note that the numerical implementation of \eqnref{eqn:full correlation estimator} has two particular difficulties. First, the integration of complex numbers is not well handled by some low-level integration routines. Secondly, and more severely, the propagators tend to Dirac delta functions of position in the limit where the time arguments are equal (or separated by an integer number of harmonic oscillator periods). That poses no problem for the second term due to the thermal, imaginary-time component, but means that numerical integrals of the first part can be non-convergent, or even worse, biased. In \figref{fig:two time two position measure}, we have not presented numerical results close to the divergences of the commutator for that reason. We note that evaluation of \figref{fig:two time two position measure} took about 80 hours of processor time.

\bibliographystyle{prsty}
\bibliography{nyman_refs}

\begin{thebibliography}{10}

\bibitem{Yasuda96}
M. Yasuda and F. Shimizu, Phys. Rev. Lett. {\bf 77},  3090  (1996).

\bibitem{Schellekens05}
M. Schellekens, R. Hoppeler, A. Perrin, J.~V. Gomes, D. Boiron, A. Aspect, and
  C.~I. Westbrook, Science {\bf 310},  648  (2005).

\bibitem{Oettl05}
A. \"Ottl, S. Ritter, M. K\"ohl, and T. Esslinger, Phys. Rev. Lett. {\bf 95},
  090404  (2005).

\bibitem{Jeltes07}
T. Jeltes, J. McNamara, W. Hogervorst, W. Vassen, V. Krachmalnicoff, M.
  Schellekens, A. Perrin, H. Chang, D. Boiron, A. Aspect, and C. Westbrook,
  Nature {\bf 445},  402  (2007).

\bibitem{Guarrera11}
V. Guarrera, P. W\"urtz, A. Ewerbeck, A. Vogler, G. Barontini, and H. Ott,
  Phys. Rev. Lett. {\bf 107},  160403  (2011).

\bibitem{Hodgman11}
S.~S. Hodgman, R.~G. Dall, A.~G. Manning, K.~G.~H. Baldwin, and A.~G. Truscott,
  Science {\bf 331},  1046  (2011).

\bibitem{Guarrera12}
V. Guarrera, D. Muth, R. Labouvie, A. Vogler, G. Barontini, M. Fleischhauer,
  and H. Ott, Phys. Rev. A {\bf 86},  021601  (2012).

\bibitem{Dall13}
R. Dall, A. Manning, S. Hodgman, W. RuGway, K. Kheruntsyan, and A. Truscott,
  Nature Physics {\bf 9},  341  (2013).

\bibitem{Altman04}
E. Altman, E. Demler, and M.~D. Lukin, Phys. Rev. A {\bf 70},  013603  (2004).

\bibitem{Foelling05}
S. F{\"o}lling, F. Gerbier, A. Widera, O. Mandel, T. Gericke, and I. Bloch,
  Nature {\bf 434},  481  (2005).

\bibitem{Greiner05}
M. Greiner, C.~A. Regal, J.~T. Stewart, and D.~S. Jin, Phys. Rev. Lett. {\bf
  94},  110401  (2005).

\bibitem{Esteve06}
J. Esteve, J.-B. Trebbia, T. Schumm, A. Aspect, C.~I. Westbrook, and I.
  Bouchoule, Phys. Rev. Lett. {\bf 96},  130403  (2006).

\bibitem{Armijo10}
J. Armijo, T. Jacqmin, K.~V. Kheruntsyan, and I. Bouchoule, Phys. Rev. Lett.
  {\bf 105},  230402  (2010).

\bibitem{Mueller10}
T. M\"uller, B. Zimmermann, J. Meineke, J.-P. Brantut, T. Esslinger, and H.
  Moritz, Phys. Rev. Lett. {\bf 105},  040401  (2010).

\bibitem{Sanner10}
C. Sanner, E.~J. Su, A. Keshet, R. Gommers, Y.-i. Shin, W. Huang, and W.
  Ketterle, Phys. Rev. Lett. {\bf 105},  040402  (2010).

\bibitem{Sanner11}
C. Sanner, E.~J. Su, A. Keshet, W. Huang, J. Gillen, R. Gommers, and W.
  Ketterle, Phys. Rev. Lett. {\bf 106},  010402  (2011).

\bibitem{Meineke12}
J. Meineke, J.-P. Brantut, D. Stadler, T. M{\"u}ller, H. Moritz, and T.
  Esslinger, Nature Physics {\bf 8},  455  (2012).

\bibitem{Zhou11}
Q. Zhou and T.-L. Ho, Phys. Rev. Lett. {\bf 106},  225301  (2011).

\bibitem{Pathria}
R. Pathria, {\em Statistical Mechanics}, {\em International series in natural
  philosophy} (Butterworth-Heinemann, Oxford, 1996).

\bibitem{Naraschewski99}
M. Naraschewski and R.~J. Glauber, Phys. Rev. A {\bf 59},  4595  (1999).

\bibitem{Goldstein98}
E.~V. Goldstein, O. Zobay, and P. Meystre, Phys. Rev. A {\bf 58},  2373
  (1998).

\bibitem{Aharonov90}
Y. Aharonov and L. Vaidman, Phys. Rev. A {\bf 41},  11  (1990).

\bibitem{Kohnen11b}
M. Kohnen, P. Petrov, R. Nyman, and E. Hinds, New Journal of Physics {\bf 13},
  085006  (2011).

\bibitem{Sakurai}
J. Sakurai, {\em Modern quantum mechanics} (Addison-Wesley California, Reading,
  MA, 1985).

\bibitem{Note1}
The notation \protect \ensuremath {{\protect \mathrm {E}\left ({X}\right )}}
  for the expectation value of a random variable is fairly standard when
  distinguishing between stochastic processes (e.g. experimental realisations)
  and quantum expectation values\cite {BreuerPetruccione,WuertzThesis}. The
  interpretation of the two kinds of expectation, statistical and quantum, is
  of course rather tricky, but here we are just making use of the clarity of
  notation.

\bibitem{Nyman11}
R. Nyman, S. Scheel, and E. Hinds, Quantum Information Processing  1  (2011).

\bibitem{Barlow}
R. Barlow, {\em Statistics: a guide to the use of statistical methods in the
  physical sciences} (John Wiley \& Sons Inc, Chichester, 1989), Vol.~29.

\bibitem{Lye03}
J.~E. Lye, J.~J. Hope, and J.~D. Close, Phys. Rev. A {\bf 67},  043609  (2003).

\bibitem{Bezett14}
A. Bezett, H.~J. van Driel, M.~P. Mink, H.~T.~C. Stoof, and R.~A. Duine, Phys.
  Rev. A {\bf 89},  023632  (2014).

\bibitem{Wilson07}
S.~D. Wilson, A.~R.~R. Carvalho, J.~J. Hope, and M.~R. James, Phys. Rev. A {\bf
  76},  013610  (2007).

\bibitem{BreuerPetruccione}
H. Breuer and F. Petruccione, {\em The Theory of Open Quantum Systems} (Oxford
  University Press, Oxford, 2007).

\bibitem{WuertzThesis}
{Peter W\"urtz}, Ph.D. thesis, {Universit\"at Kaiserslautern}, 2012.

\end{thebibliography}

\end{document}